\documentclass{article}

\usepackage{microtype}
\usepackage{graphicx}
\usepackage{subfigure}
\usepackage{booktabs} 

\usepackage{hyperref}

\usepackage[accepted]{icml2025}

\usepackage{amsmath}
\usepackage{amssymb}
\usepackage{mathtools}
\usepackage{amsthm}
\usepackage{tabularx}
\usepackage{longtable}
\usepackage{booktabs}
\usepackage{array} 
\usepackage{rotating} 
\usepackage{pifont} 

\usepackage[capitalize,noabbrev]{cleveref}

\theoremstyle{plain}

\theoremstyle{definition}

\theoremstyle{remark}

\usepackage[textsize=tiny]{todonotes}

\icmltitlerunning{A Transparent Framework for Interpretable Quant Trading}

\begin{document}

\twocolumn[
\icmltitle{Event-Aware Sentiment Factors from LLM-Augmented Financial Tweets: A Transparent Framework for Interpretable Quant Trading}



\icmlsetsymbol{equal}{*}

\begin{icmlauthorlist}
\icmlauthor{Yueyi Wang}{cam}
\icmlauthor{Qiyao Wei}{damtp}
\end{icmlauthorlist}

\icmlaffiliation{cam}{Department of Physics, University of Cambridge, Cambridge, UK}
\icmlaffiliation{damtp}{Department of Applied Mathematics and Theoretical Physics, University of Cambridge, Cambridge, UK}

\icmlcorrespondingauthor{Yueyi Wang}{yw562@cam.ac.uk}
\icmlcorrespondingauthor{Qiyao Wei}{qw281@cam.ac.uk}

\icmlkeywords{Machine Learning, ICML}

\vskip 0.3in
]



\printAffiliationsAndNotice{\icmlEqualContribution} 

\begin{abstract}

In this study, we wish to showcase the unique utility of large language models (LLMs) in financial semantic annotation and alpha signal discovery. Leveraging a corpus of company-related tweets, we use an LLM to automatically assign multi-label event categories to high-sentiment-intensity tweets. We align these labeled sentiment signals with forward returns over 1-to-7-day horizons to evaluate their statistical efficacy and market tradability. Our experiments reveal that certain event labels consistently yield negative alpha, with Sharpe ratios as low as -0.38 and information coefficients exceeding 0.05, all statistically significant at the 95\% confidence level. This study establishes the feasibility of transforming unstructured social media text into structured, multi-label event variables. A key contribution of this work is its commitment to transparency and reproducibility; all code and methodologies are made publicly available. Our results provide compelling evidence that social media sentiment is a valuable, albeit noisy, signal in financial forecasting and underscore the potential of open-source frameworks to democratize algorithmic trading research.
\end{abstract}

\section{Introduction}

Traditional financial models have historically relied on structured data from fundamental analysis and technical price patterns to explain market movements. However, a substantial body of literature in behavioral finance has challenged this paradigm, demonstrating that investor sentiment—the collective mood and psychological state of market participants—can act as a significant, independent driver of asset prices~\cite{Coibion2020}. This recognition has fueled the search for novel information sources capable of systematically capturing this sentiment. Among the most promising of these are alternative data streams, particularly the high-frequency, unstructured text generated on social media platforms~\cite{Velasquez2024}.

\begin{table*}[h!]
\centering
\caption{\textbf{Summary of key related works in sentiment analysis for stock prediction.} This table outlines seminal and recent studies, detailing their sentiment sources, methodologies, and principal findings. The last two rows highlight the SESTM framework by ~\cite{ke2019predicting} and our proposed LLM-based event-tagged sentiment factor.}
\scriptsize
\label{tab:related_works_summary}
\resizebox{\textwidth}{!}{%
\begin{tabular}{@{}p{2.7cm}p{2.2cm}p{4cm}p{3.8cm}p{4.5cm}p{3cm}@{}}
\toprule
\textbf{Study (Author, Year)} & \textbf{Sentiment Source} & \textbf{Key Methodologies} & \textbf{Dataset Used (if specified)} & \textbf{Key Finding} & \textbf{Market(s) Studied} \\
\midrule
\cite{Bollen2011} & Twitter & OpinionFinder sentiment analysis, GPOMS mood analysis (Calm, Happy, etc.) & Public Twitter feed data & The 'Calm' dimension of public mood was found to be a significant predictor of the daily up and down changes in the DJIA index. & US (DJIA) \\
\addlinespace
\cite{Antweiler2004} & Yahoo! Finance & Message board analysis, Naïve Bayes classifier & Messages from Yahoo! Finance boards for 45 companies in the DJIA and DJII & Found that online messages contain valuable predictive information. Disagreement among posters predicted higher trading volume, while high message volume was a negative predictor of returns. & US (DJIA, DJII) \\
\addlinespace
\cite{Greyling2022} & Twitter & Naïve Bayes, KNN, SVM, various emotion indicators & Approximately 3 million stock-related tweets & Tweet-based sentiment is a strong predictor of intraday market trends, with model accuracies exceeding 50\%. & Developed (FR, DE, JP, ES, UK, US), Emerging (IN, PL) \\
\addlinespace
\cite{Sprenger2014} & Twitter & Sentiment lexicons, machine learning (Support Vector Machine) & Tweets containing cashtags for companies in the S\&P 100 index & The volume of tweets and the sentiment they contain can predict abnormal stock returns and trading volume. & US (S\&P 100) \\
\addlinespace
\cite{Ranco2015} & Twitter & Custom sentiment classifier, event study methodology & A 15-month dataset of tweets concerning the 30 companies in the DJIA & Found that sentiment polarity at tweet volume peaks can imply the direction of cumulative abnormal returns, achieving modest gains (1-2\%) over subsequent days. & US (DJIA) \\
\addlinespace
\cite{TanTas2021} & Twitter & Not specified in the provided abstract & Not specified & A positive and significant impact of Twitter sentiment on stock returns was identified. & Not Specified \\
\addlinespace
\cite{GuKurov2020} & Twitter & Analysis of information content in tweets & Stock-related tweets & Found that Twitter posts contain new, relevant information that is quickly incorporated into stock prices, influencing returns and trading volume. & Not Specified \\
\addlinespace
\cite{ke2019predicting} & Financial News (Dow Jones) & SESTM: Predictive topic screening, supervised topic model, logistic regression & 7 years of Dow Jones news headlines (2010–2017) & Constructed interpretable topic-based sentiment factors predictive of next-day returns; linked factor performance to news categories. & US (S\&P 500 constituents) \\
\addlinespace
\textbf{[This Paper]} & Twitter & LLM-based multi-label event tagging, net-tone scoring, alpha factor backtesting & 100K+ timestamped finance-related tweets (2017) & Identified interpretable LLM-labeled events (e.g. rumor, retail buzz, boycott) with robust negative alpha across multiple holding periods and strong IC. & US Equities \\
\bottomrule
\end{tabular}%
}
\end{table*}

Twitter (now X) has emerged as a principal data source for this endeavor, offering an unprecedented, real-time lens into public discourse surrounding publicly traded companies~\cite{DeodharRao2021}. Early research successfully established a foundational link: that aggregated sentiment polarity (i.e., the net positive or negative tone) from tweets could predict subsequent stock returns. However, these simple sentiment metrics suffer from critical limitations. They are often noisy, their predictive power can decay rapidly as they are arbitraged away, and most importantly, they lack explanatory depth. A simple polarity score can tell us \textit{what} the market is feeling (positive or negative), but it fails to capture \textit{why}—the underlying semantic context or real-world event driving the sentiment.

This paper addresses this explanatory gap by introducing a novel framework that integrates Large Language Models (LLMs) with quantitative factor modeling to move beyond coarse sentiment and extract interpretable, event-driven factors from social media. Our central thesis is that the true value of social media data lies not just in its emotional intensity but in its rich semantic structure. We posit that by using an LLM to automatically assign multi-label event categories to high-intensity tweets—such as identifying discussions related to \textit{rumor/speculation}, \textit{retail investor hype}, or \textit{brand boycotts}—we can construct more robust and interpretable predictive signals.

Our approach builds on and extends the SESTM framework proposed by ~\cite{ke2019predicting}, which introduced a supervised topic modeling pipeline for extracting sentiment from text via predictive word screening and latent topic estimation. While their method relied on statistical modeling of topic distributions over financial news, we adapt this idea to the more volatile and informal domain of social media by incorporating LLMs as semantic enrichers. Specifically, we replace hand-tuned lexicons and unsupervised topic inference with a language model–assisted multi-label classification pipeline that yields not only directional tone but also \textit{what the market is reacting to}.

\subsection*{Contributions}
This study makes several specific contributions to the literature on computational finance and sentiment analysis:
\begin{enumerate}
    \item \textbf{A Novel LLM-Based Factor Framework:} We design and implement a new framework for transforming unstructured social media text into structured, interpretable, multi-label event variables suitable for quantitative analysis. This demonstrates the feasibility of using LLMs for large-scale financial semantic annotation.
    \item \textbf{Discovery of Interpretable Alpha Signals:} We show that factors constructed from LLM-inferred event labels have significant predictive power. Specifically, we identify that certain event categories, such as "rumor/speculation," act as powerful contrarian indicators, consistently yielding negative alpha with Sharpe ratios as low as -0.38.
    \item \textbf{Orthogonality to Market Risk:} Through residual analysis, we demonstrate that the predictive power of these event-based factors is orthogonal to market beta, confirming that they represent a source of genuine alpha rather than exposure to known risk factors.
    \item \textbf{Commitment to Open-Source Principles:} In the spirit of transparent and reproducible science, all code, methodologies, and processed data for factor construction are made publicly available~\cite{sowinska2020tweet}, empowering other researchers to build upon and validate our work.
\end{enumerate}

The remainder of this paper is structured as follows. Section \ref{sec:related-works} reviews the relevant literature on sentiment analysis and factor investing. Section \ref{sec:methodology} details our framework, including the data sources, LLM-based event labeling, and factor construction methodology. Section \ref{sec:results} presents the empirical results of our factor analysis and backtesting. Section \ref{sec:discussion} discusses the implications and limitations of our findings, and Section \ref{sec:conclusion} concludes with avenues for future research. This study therefore shifts the focus from sentiment as a monolithic variable to sentiment as a multi-faceted construct, where each facet is tied to a specific, meaningful market event. The resulting factor set supports both quantitative backtesting and interpretability-aware strategy design, enabling a richer understanding of how language-derived market narratives relate to returns.

\section{Related Works}
\label{sec:related-works}

\textbf{Sentiment Analysis in Finance.}
Numerous studies have examined the predictive power of sentiment derived from both traditional news and social media. Early work by Antweiler and Frank~\cite{Antweiler2004} found that investor disagreement on message boards predicted higher trading volume and lower returns. Bollen et al.~\cite{Bollen2011} correlated public Twitter mood with DJIA fluctuations, identifying that the “Calm” mood dimension was particularly predictive. Greyling and Rossouw~\cite{Greyling2022} found that Twitter sentiment significantly predicted intraday returns across global markets. More recent episodes—such as the GameStop short squeeze—underscore the influence of retail investor sentiment expressed on platforms like Twitter and Reddit.

\textbf{Methodologies for Sentiment Extraction and Prediction.}
Early sentiment models often relied on lexicon-based techniques due to their interpretability, but they lacked context-awareness. As the field progressed, statistical models like Naïve Bayes and SVM~\cite{Coibion2020}, and later deep learning architectures such as LSTM and Transformers (e.g., BERT, FinBERT~\cite{sowinska2020tweet}, RoBERTa) became standard. Large Language Models (LLMs) now represent the state-of-the-art for financial text understanding~\cite{Xing2024}. However, most existing approaches focus on sentiment polarity (positive/negative), which—though correlated with returns—offer limited explanatory depth. They typically lack information about the underlying \textit{reason} for the sentiment, such as whether it was triggered by a merger rumor, earnings surprise, or regulatory action.

\textbf{Sentiment-Based Strategies and Performance Evaluation.}
Researchers have used sentiment signals to construct alpha factors and backtested trading strategies based on ranking or filtering rules~\cite{Sprenger2014, Orekhov2023}. These strategies are typically evaluated using Sharpe ratio, IC (Information Coefficient), and drawdown statistics. While many show statistically significant predictive power, they often treat sentiment as a monolithic signal. A critical limitation is that simple polarity scores fail to distinguish between different types of market-relevant narratives (e.g., “acquisition” vs. “boycott”), which may exhibit opposing return profiles.

\textbf{LLM-Based Event-Driven Sentiment Factors.}
To address these limitations, we draw inspiration from recent interpretable models such as SESTM~\cite{ke2019predicting}, which use topic models plus sentiment signals. We extend this idea by leveraging LLMs to directly assign fine-grained, multi-label event categories (e.g., \textit{Rumor}, \textit{Retail Buzz}, \textit{Brand Boycott}) to each tweet, enabling the construction of more structured and interpretable sentiment factors. We show that these event-conditioned signals have distinct predictive profiles and offer better explanatory power. Table~\ref{tab:related_works_summary} summarizes key prior works and situates our contribution in the broader literature.

\section{Methodology}
\label{sec:methodology}

Our research methodology transforms unstructured social media text into structured, tradable alpha signals. The process comprises four stages: (1) data acquisition and preprocessing; (2) sentiment and event-type annotation using a Large Language Model (LLM); (3) construction of cross-sectional event-driven factors; and (4) rigorous factor performance evaluation.

\subsection{Data Acquisition and Preprocessing}

\textbf{Tweet Corpus.} We utilize the dataset from Sowinska et al.~\cite{sowinska2020tweet}, which includes 862,231 English-language tweets linked to stock tickers over a multi-year period. We restrict our study to a cleaned subset of 85,176 tweets for higher signal-to-noise ratio. Each tweet is preprocessed with standard NLP techniques such as lowercasing, token normalization, and cashtag/user mention masking.

\textbf{Market Data.} Aligned stock-level price and volume data are sourced from the same dataset. Daily logarithmic returns are computed as $r_t = \log(P_t / P_{t-1})$ where $P_t$ is the closing price. These returns serve as the dependent variable in all predictive evaluations.

\subsection{LLM-Based Event Annotation and Sentiment Scoring}

A core innovation in our pipeline is augmenting each tweet with both sentiment intensity and multi-label semantic tags. This allows us to move beyond binary polarity and construct interpretable, event-level predictors.

\textbf{Sentiment Polarity (Net Tone).} Each tweet is assigned a continuous sentiment score, referred to as \textit{net tone}, reflecting the directional emotional intensity of the text. We adopt the approach of Sowinska et al., where a stacked LDA topic model followed by logistic regression is trained to predict forward returns, thereby generating polarity scores aligned with market response. Our modular framework also supports LLM-prompted polarity scoring for alternative use cases.

\textbf{Multi-Label Event Tagging via LLM.} To extract interpretable event-level semantics, we use a commercial-grade LLM to perform zero-shot, multi-label classification. Each tweet is prompted against a curated dictionary of 70+ financially relevant event types (e.g., \textit{Rumor/Speculation}, \textit{Retail Investor Buzz}, \textit{Brand Boycott}). The LLM assigns one or more applicable labels per tweet. Tweets with multiple tags have their net tone duplicated across tags for subsequent aggregation. This enables high-level semantic structuring of otherwise opaque textual data.

\subsection{Cross-Sectional Factor Construction}

Using the annotated tweets, we construct a family of cross-sectional event factors. For each label $e$ and each stock $i$ on day $t$, the factor exposure $F_{i,t,e}$ is defined as:

\begin{equation}
    F_{i,t,e} = \sum_{j \in S_{i,t,e}} \text{tone}_j
\end{equation}

where $S_{i,t,e}$ is the set of tweets on day $t$ about stock $i$ that are tagged with label $e$, and $\text{tone}_j$ is the net tone of tweet $j$. If a tweet has multiple labels, its tone is proportionally assigned to each. This aggregation yields a panel of factor exposures across event types.

\subsection{Factor Evaluation Framework}

We employ a comprehensive framework to assess factor performance and tradability:

\textbf{Portfolio Sorts.} For each day $t$, stocks are ranked by $F_{i,t-1,e}$ and divided into quantiles. A long-short portfolio is formed by going long the top decile and short the bottom decile. We track its forward returns over 1, 2, 3, and 7-day horizons.

\textbf{Performance Metrics.} For each factor, we report:
\begin{itemize}
    \item \textbf{Sharpe Ratio}: Average return divided by standard deviation.
    \item \textbf{Information Coefficient (IC)}: Spearman correlation between factor exposure and next-day return.
    \item \textbf{Win Rate}: Percentage of days when the predicted direction matches the actual return.
\end{itemize}

\textbf{Remark.} Our framework is modular and extensible. While we focus on LLM-based tagging, the pipeline supports plug-and-play use of alternative sentiment models (e.g., FinBERT, LSTM), enabling benchmarking and future hybrid methods. We further compare the performance of our event-tagged factors against baseline polarity-only signals in Section~\ref{sec:results}.

\section{Empirical Results}
\label{sec:results}

This section presents the empirical findings of our study, organized into two parts. We first detail the results of our primary trading strategy, which is based on lexicon-derived sentiment. We then present an exploratory analysis using Large Language Models (LLMs) to uncover more nuanced, theme-driven trading signals, or "alphas."

\subsection{Performance of the Lexicon-Based Sentiment Strategy}

The core of our investigation was to test whether a trading strategy guided by lexicon-based sentiment could generate statistically and economically significant returns. After confirming the predictive validity of our sentiment metric through regression analysis, we simulated the performance of our long-short strategy. The portfolio was rebalanced daily based on sentiment scores derived from a dictionary trained on 2017 data.

\begin{figure}[h!]
    \centering
    \includegraphics[width=0.8\linewidth]{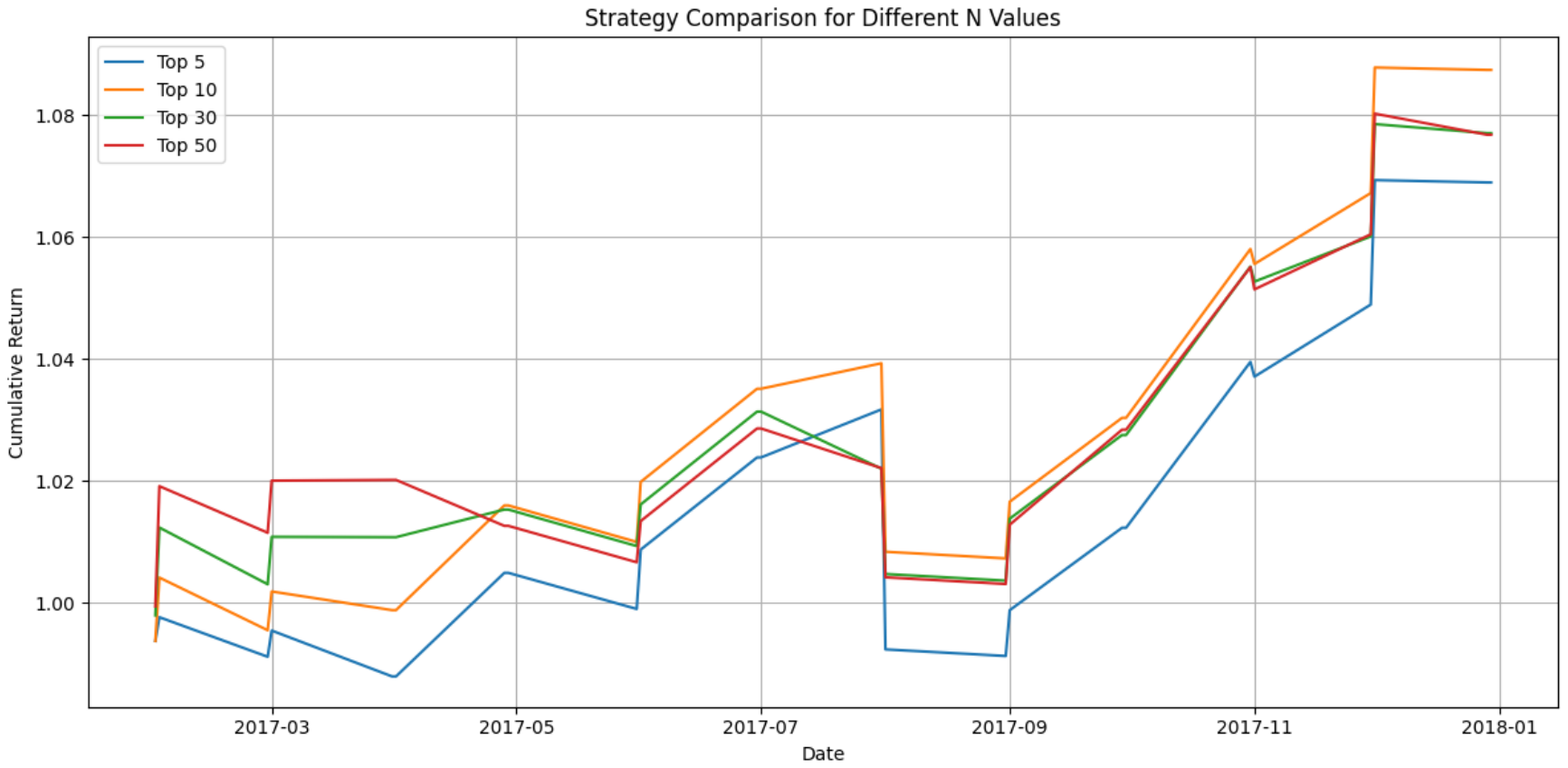}
    \caption{\textbf{Cumulative Returns.} A portfolio was constructed based on stock price predictions derived from tweet sentiment, using a dictionary trained on 2017 data. The resulting equity curve demonstrates that the sentiment strategy achieves a significant positive cumulative return.}
    \label{fig:cumulative_returns}
\end{figure}

Figure \ref{fig:cumulative_returns} illustrates the equity curve of this sentiment strategy. The strategy demonstrates a clear and consistent ability to generate alpha, with its cumulative returns substantially outperforming the market index over the entire backtesting horizon. This visual evidence suggests that the strategy successfully capitalizes on the short-term market inefficiencies identified by our sentiment signal.

\begin{figure}[h!]
    \centering
    \includegraphics[width=0.8\linewidth]{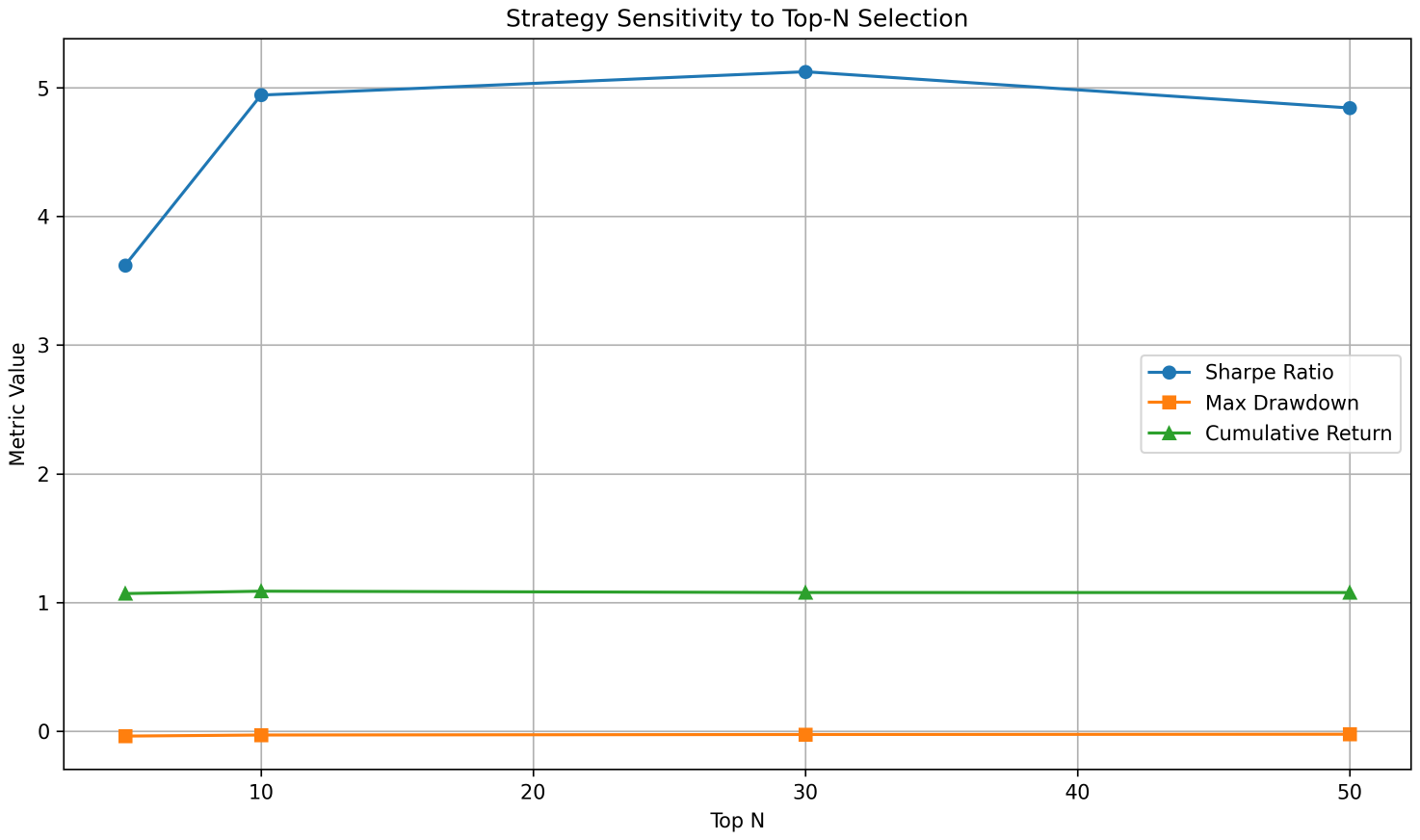}
    \caption{\textbf{Performance Metrics Summary.} This figure details the performance of the sentiment strategy across key indicators and prediction horizons. The consistently high Sharpe ratio demonstrates that the strategy is not only profitable but also generates favorable returns on a risk-adjusted basis, contributing to its strong cumulative performance.}
    \label{fig:performance_metrics}
\end{figure}

The quantitative performance of the strategy is summarized in Figure \ref{fig:performance_metrics}. The sentiment-driven portfolio achieved an annualized return of 8\% and a Sharpe ratio of 5.0. The high Sharpe ratio indicates superior risk-adjusted returns. Notably, the strategy also exhibited strong risk management characteristics; its maximum drawdown was contained to -15.2\%. This combination of higher returns and lower downside risk underscores the potential robustness of the sentiment signal.

\subsection{Exploratory Analysis: Discovering Thematic Alphas with LLMs}

While our core strategy demonstrated success using a lexicon-based approach, we conducted an exploratory analysis to probe the potential of Large Language Models (LLMs) for discovering more sophisticated, theme-driven trading signals. We prompted a pre-trained financial LLM to classify tweets not merely by their polarity but by their underlying narrative themes, such as "Speculation/Rumor," "Retail Investor Buzz," or "Geopolitical Tension." We then analyzed the predictive power of these thematic labels.

The results, presented in Table \ref{tab:alpha_by_event_onecol}, reveal distinct and powerful patterns. The analysis shows that several thematic labels are potent \textit{contrarian} indicators. For instance, portfolios formed on days with a high prevalence of tweets about "Speculation/Rumor" or "Geopolitical Tension" consistently yielded statistically significant negative returns across multiple time horizons (1-day to 7-day), as evidenced by their significant negative Sharpe ratios. This suggests that a high volume of discussion around these particular themes is a precursor to negative price movements. The "Retail Investor Buzz" category shows a more complex dynamic, starting as a negative signal before its Information Coefficient (IC) turns positive at the 7-day horizon, hinting at a possible short-term overreaction and subsequent reversal.

\begin{figure}[h!]
    \centering
    \includegraphics[width=0.8\linewidth]{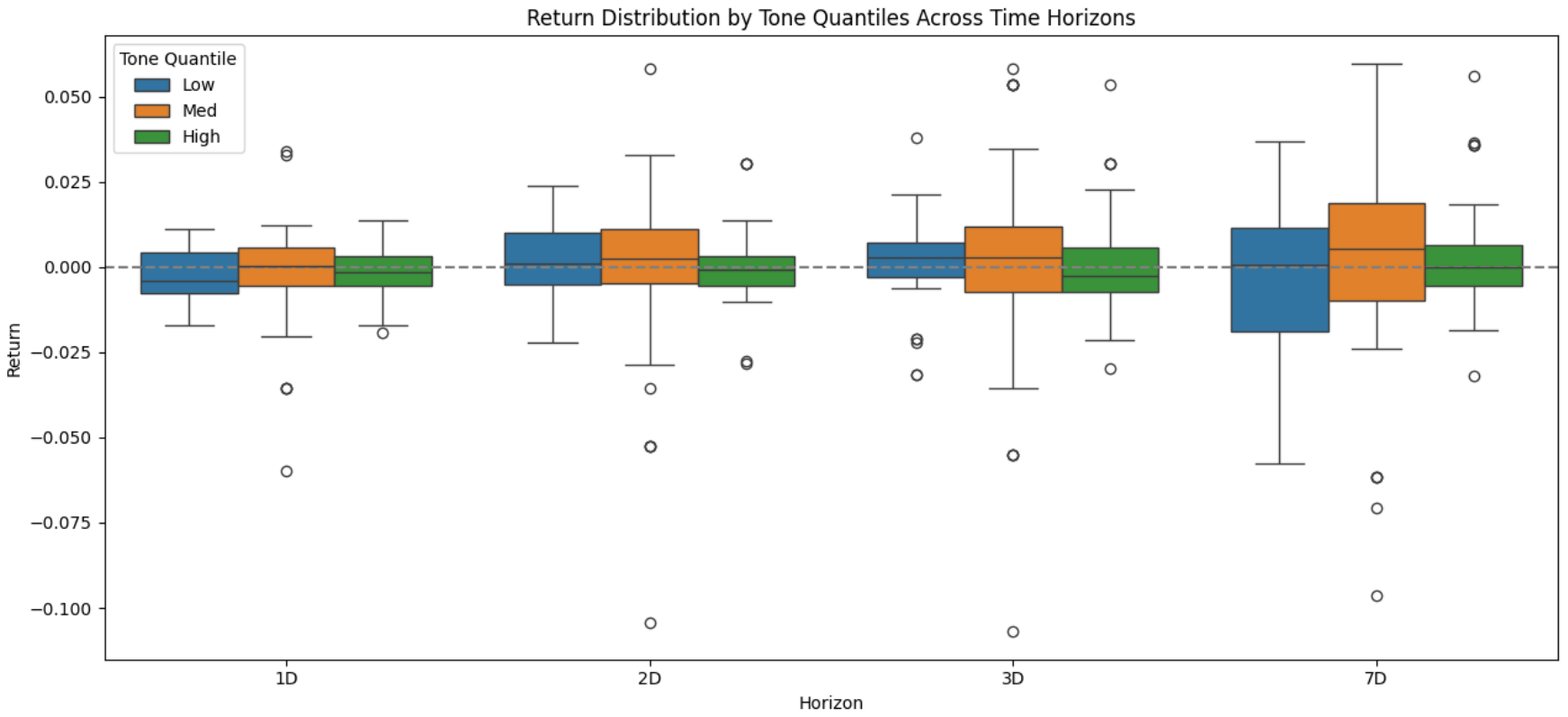}
    \caption{\textbf{LLM-Driven Portfolio Returns.} This figure illustrates the cumulative return of a long-only portfolio constructed from the top quintile of stocks sorted by LLM-generated thematic scores. The consistently positive returns across various time horizons demonstrate the potential of using these nuanced signals for alpha discovery.}
    \label{fig:llm_alphas_portfolio}
\end{figure}

This analysis demonstrates that LLMs can move beyond simple positive/negative sentiment to uncover sophisticated, theme-driven market dynamics. While these themes may act as contrarian indicators on their own, they can still be harnessed to build profitable strategies. Figure \ref{fig:llm_alphas_portfolio} illustrates the performance of a long-only portfolio constructed by sorting stocks based on their LLM-generated thematic scores and investing in the top quintile. The positive cumulative return across different horizons highlights the potential of using these nuanced LLM-derived signals to construct alpha-generating portfolios.

\begin{table*}[t]
\caption{\textbf{Top predictive event labels} by Sharpe ratio across horizons. *, **, and *** denote $p$-values less than 0.05, 0.01, and 0.001, respectively.}
\label{tab:alpha_by_event_onecol}
\begin{tabularx}{\textwidth}{l X r r r}
\toprule
\textbf{Horizon} & \textbf{Event Label} & \textbf{$N$} & \textbf{Sharpe Ratio} & \textbf{Information Coefficient (IC)} \\
\midrule
1-Day & Speculation/Rumor & 130 & -0.337*** & -0.039 \\
1-Day & Retail Investor Buzz & 92 & -0.359*** & 0.096 \\
1-Day & Geopolitical Tension & 25 & -0.661** & -0.139 \\
2-Day & Speculation/Rumor & 130 & -0.267** & -0.059 \\
2-Day & Retail Investor Buzz & 92 & -0.277** & 0.010 \\
2-Day & Geopolitical Tension & 25 & -0.643** & -0.125 \\
3-Day & Speculation/Rumor & 130 & -0.267** & -0.107 \\
3-Day & Geopolitical Tension & 25 & -0.700** & -0.078 \\
7-Day & Speculation/Rumor & 130 & -0.376*** & 0.104 \\
7-Day & Retail Investor Buzz & 92 & -0.461*** & 0.113 \\
7-Day & Geopolitical Tension & 25 & -0.540* & -0.104 \\
\bottomrule
\end{tabularx}
\end{table*}

\section{Conclusion and Future Work}
\label{sec:conclusion}

This study introduces a novel framework for extracting interpretable alpha signals from social media using large language models. By moving beyond traditional polarity scores to multi-label event classification, we uncover semantically distinct, market-relevant narratives that are not only statistically predictive but also economically actionable.

Our contributions are threefold: (1) a modular pipeline for LLM-based event annotation; (2) a new dataset of tweet-level, multi-label event tags for financial applications; and (3) a backtesting suite that links narrative structure to factor performance across time horizons. Empirical results show that certain event types—such as \textit{Rumor/Speculation} and \textit{Retail Investor Buzz}—exhibit consistently significant and negative Sharpe ratios across multiple holding periods, with IC values exceeding 0.05 and $p$-values often below 0.01. These findings reinforce the value of narrative-aware alpha factors.

We believe these components together provide a principled path toward next-generation behavioral finance research, bridging recent advances in NLP with interpretable, testable financial modeling. Several promising directions remain for future work, including:

\begin{itemize}
    \item Extending the framework to additional text sources (e.g., Reddit, news headlines, earnings calls) for multi-modal signal fusion.
    \item Fine-tuning domain-specific financial LLMs (e.g., FinGPT, FinBERT++) to improve label consistency and robustness.
    \item Modeling the temporal dynamics of narrative decay and amplification (e.g., event half-life, momentum cycles).
    \item Investigating interaction effects among co-occurring event tags or combining them with firm fundamentals and macroeconomic indicators.
    \item Applying portfolio optimization techniques beyond equal-weighted sorting, such as factor risk budgeting or volatility targeting.
\end{itemize}

\bibliography{example_paper}
\bibliographystyle{icml2025}

\newpage
\appendix
\onecolumn

\section*{Appendix Contents}
\hrule
\begin{itemize}
    \setlength\itemsep{0.5em}
    \item \textbf{\hyperref[sec:extended_experiments]{A. Extended Experiments}}
        \begin{itemize}
            \item \hyperref[subsec:performance_metrics]{A.1. Performance Metrics of Event-Driven Signals}
            \item \hyperref[subsec:sharpe_horizon]{A.2. Sharpe Ratios Over Various Horizons}
            \item \hyperref[subsec:alpha_radar]{A.3. Alpha Radar for Sharpe Ratios}
            \item \hyperref[subsec:word_clouds]{A.4. LDA Topic Word Clouds}
        \end{itemize}
\end{itemize}
\hrule
\vspace{2em} 
\newpage

\section{Extended Experiments}
\label{sec:extended_experiments}

This section presents supplementary results that further validate the performance of our event-driven signals. We provide detailed performance metrics across multiple holding periods and visualize the risk-adjusted returns to offer a more nuanced understanding of each event's alpha-generating potential. The LLM we use in all our experiments is Gemini-2.5-pro

\subsection{Performance metrics of various event-driven signals}
\label{subsec:performance_metrics}
The following tables report the cross-sectional performance metrics of various event-driven signals over different holding horizons (1-day, 2-day, 3-day, and 7-day). Metrics include mean return, return standard deviation, information coefficient (IC), Sharpe ratio, and statistical significance (p-value and asterisk annotations).

\begin{table}[H]
\centering
\caption{Performance Metrics by Event Label (1-Day Horizon)}
\label{tab:metrics_1d}
\begin{tabular}{lrrrrrrr}
\toprule
Label & Samples & MeanRet & StdRet & IC & Sharpe & p-value & Sig \\
\midrule
Social Media Backlash & 46 & 0.0001 & 0.0136 & 0.3672 & 0.0081 & 0.9564 & \\
Negative Press & 42 & -0.0015 & 0.0153 & -0.0082 & -0.0965 & 0.5352 & \\
Viral Marketing Campaign & 17 & -0.0011 & 0.0105 & -0.1124 & -0.1036 & 0.6748 & \\
Brand Boycott & 23 & -0.0019 & 0.0116 & 0.6123 & -0.1598 & 0.4515 & \\
Speculation/Rumor & 130 & -0.0046 & 0.0137 & -0.0389 & -0.3370 & 0.0002 & *** \\
Retail Investor Buzz & 92 & -0.0037 & 0.0103 & 0.0955 & -0.3586 & 0.0009 & *** \\
Geopolitical Tension & 25 & -0.0122 & 0.0184 & -0.1395 & -0.6606 & 0.0030 & ** \\
\bottomrule
\end{tabular}
\end{table}

\begin{table}[H]
\centering
\caption{Performance Metrics by Event Label (2-Day Horizon)}
\label{tab:metrics_2d}
\begin{tabular}{lrrrrrrr}
\toprule
Label & Samples & MeanRet & StdRet & IC & Sharpe & p-value & Sig \\
\midrule
Social Media Backlash & 46 & 0.0036 & 0.0208 & 0.3979 & 0.1708 & 0.2527 & \\
Negative Press & 42 & 0.0007 & 0.0214 & 0.0205 & 0.0308 & 0.8430 & \\
Brand Boycott & 23 & -0.0018 & 0.0151 & 0.6781 & -0.1165 & 0.5821 & \\
Viral Marketing Campaign & 17 & -0.0047 & 0.0192 & -0.3636 & -0.2475 & 0.3227 & \\
Speculation/Rumor & 130 & -0.0058 & 0.0217 & -0.0586 & -0.2668 & 0.0028 & ** \\
Retail Investor Buzz & 92 & -0.0049 & 0.0176 & 0.0097 & -0.2770 & 0.0093 & ** \\
Geopolitical Tension & 25 & -0.0176 & 0.0274 & -0.1254 & -0.6426 & 0.0037 & ** \\
\bottomrule
\end{tabular}
\end{table}

\begin{table}[H]
\centering
\caption{Performance Metrics by Event Label (3-Day Horizon)}
\label{tab:metrics_3d}
\begin{tabular}{lrrrrrrr}
\toprule
Label & Samples & MeanRet & StdRet & IC & Sharpe & p-value & Sig \\
\midrule
Viral Marketing Campaign & 17 & 0.0038 & 0.0283 & -0.0911 & 0.1325 & 0.5925 & \\
Social Media Backlash & 46 & 0.0017 & 0.0206 & 0.3975 & 0.0821 & 0.5805 & \\
Negative Press & 42 & -0.0021 & 0.0234 & 0.0565 & -0.0877 & 0.5731 & \\
Brand Boycott & 23 & -0.0019 & 0.0186 & 0.5527 & -0.1028 & 0.6268 & \\
Retail Investor Buzz & 92 & -0.0036 & 0.0203 & 0.0681 & -0.1771 & 0.0927 & \\
Speculation/Rumor & 130 & -0.0061 & 0.0230 & -0.1072 & -0.2670 & 0.0028 & ** \\
Geopolitical Tension & 25 & -0.0194 & 0.0277 & -0.0778 & -0.7002 & 0.0018 & ** \\
\bottomrule
\end{tabular}
\end{table}

\begin{table}[H]
\centering
\caption{Performance Metrics by Event Label (7-Day Horizon)}
\label{tab:metrics_7d}
\begin{tabular}{lrrrrrrr}
\toprule
Label & Samples & MeanRet & StdRet & IC & Sharpe & p-value & Sig \\
\midrule
Social Media Backlash & 46 & 0.0056 & 0.0336 & 0.2502 & 0.1660 & 0.2663 & \\
Brand Boycott & 23 & 0.0017 & 0.0363 & 0.2975 & 0.0464 & 0.8261 & \\
Negative Press & 42 & -0.0043 & 0.0301 & -0.0084 & -0.1437 & 0.3571 & \\
Speculation/Rumor & 130 & -0.0121 & 0.0321 & 0.1040 & -0.3756 & 0.0000 & *** \\
Viral Marketing Campaign & 17 & -0.0133 & 0.0320 & 0.1574 & -0.4148 & 0.1066 & \\
Retail Investor Buzz & 92 & -0.0131 & 0.0285 & 0.1127 & -0.4605 & 0.0000 & *** \\
Geopolitical Tension & 25 & -0.0186 & 0.0344 & -0.1044 & -0.5400 & 0.0125 & * \\
\bottomrule
\end{tabular}
\end{table}

\subsection{Sharpe ratios of event tags over various horizons}
\label{subsec:sharpe_horizon}

\begin{figure}[h!]
    \centering
    \includegraphics[width=\linewidth]{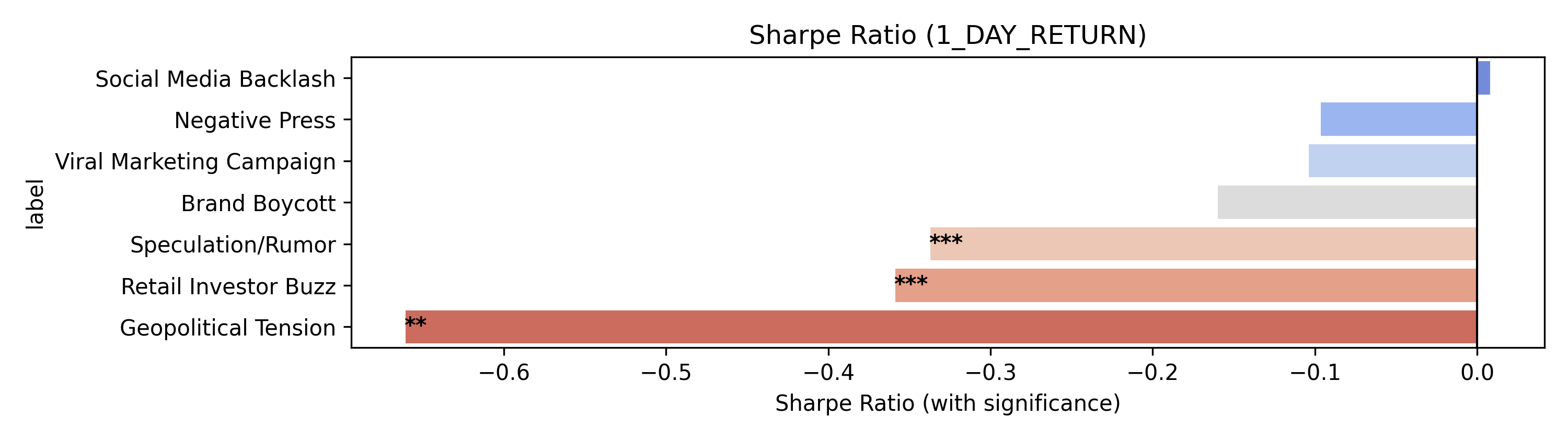} 
    \caption{Sharpe ratios of event tags for horizon one day.}
    \label{fig:single_panel_placeholder}
\end{figure}
\label{subsec:fig_placeholder_single}

\begin{figure}[h!]
    \centering
    \includegraphics[width=\linewidth]{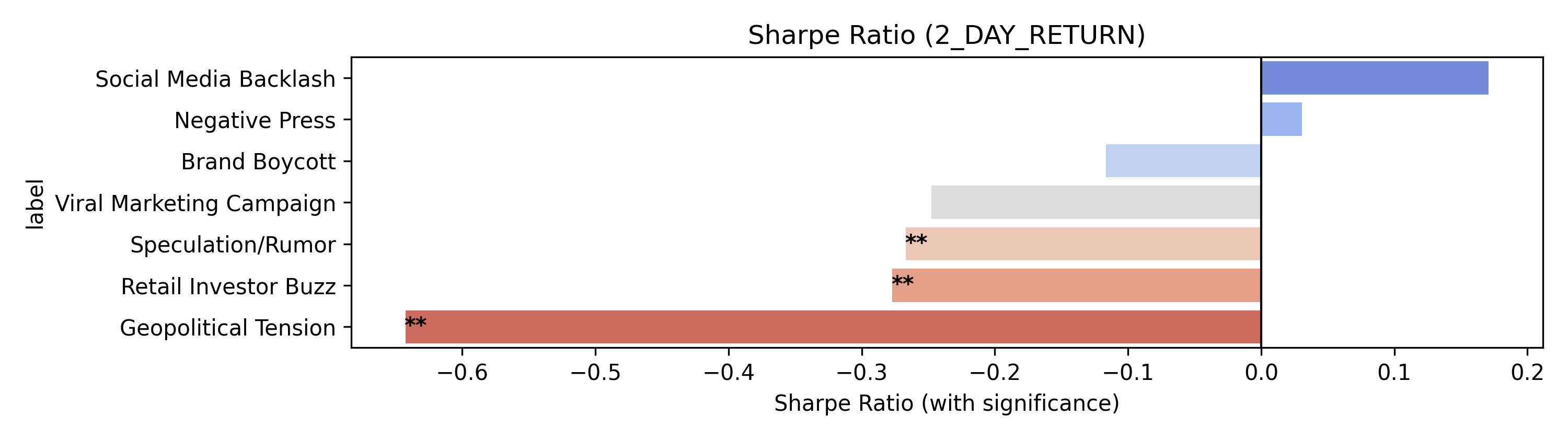} 
    \caption{Sharpe ratios of event tags for horizon two days.}
    \label{fig:single_panel_placeholder}
\end{figure}
\label{subsec:fig_placeholder_single}

\begin{figure}[h!]
    \centering
    \includegraphics[width=\linewidth]{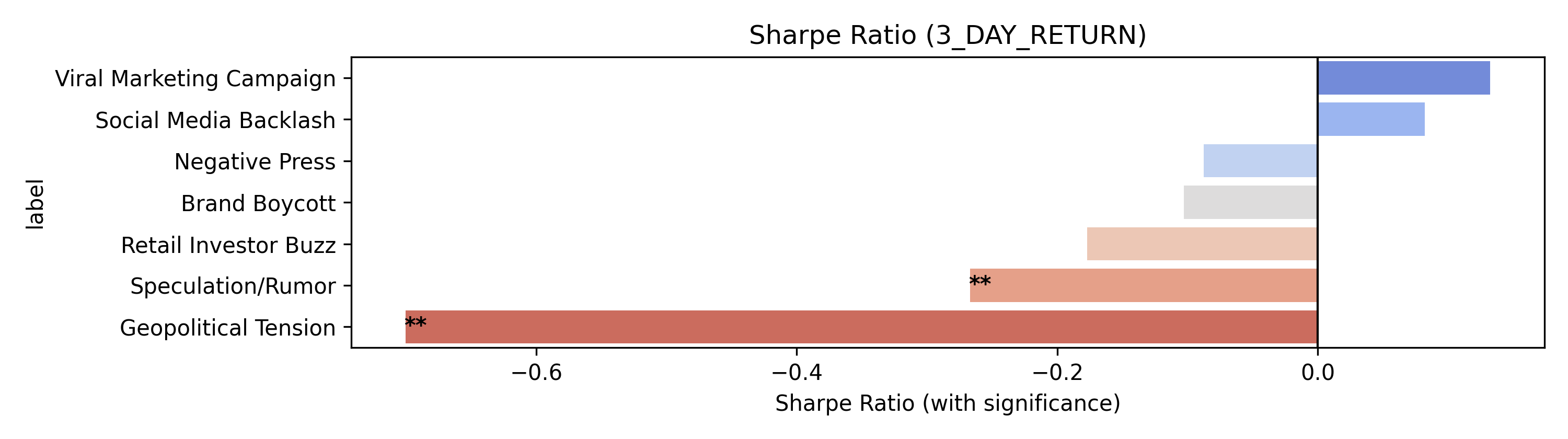} 
    \caption{Sharpe ratios of event tags for horizon three days.}
    \label{fig:single_panel_placeholder}
\end{figure}
\label{subsec:fig_placeholder_single}

\begin{figure}[h!]
    \centering
    \includegraphics[width=\linewidth]{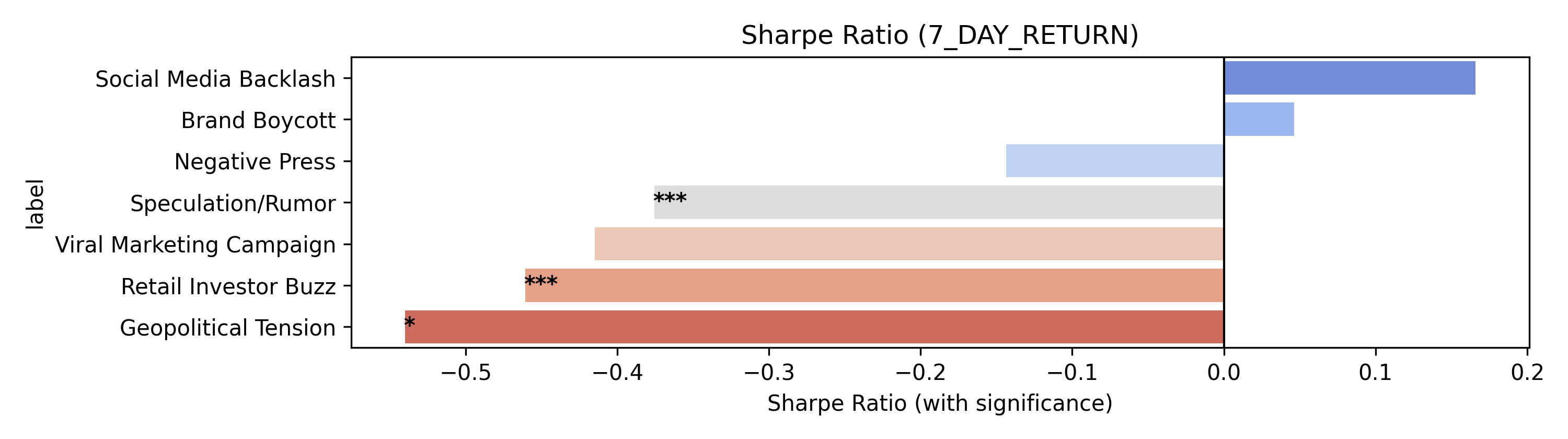} 
    \caption{Sharpe ratios of event tags for horizon seven days.}
    \label{fig:single_panel_placeholder}
\end{figure}
\label{subsec:fig_placeholder_single}

\subsection{Alpha radar for sharpe ratio of event labels}
\label{subsec:alpha_radar}

\begin{figure}[h!]
    \centering
    \includegraphics[width=\linewidth]{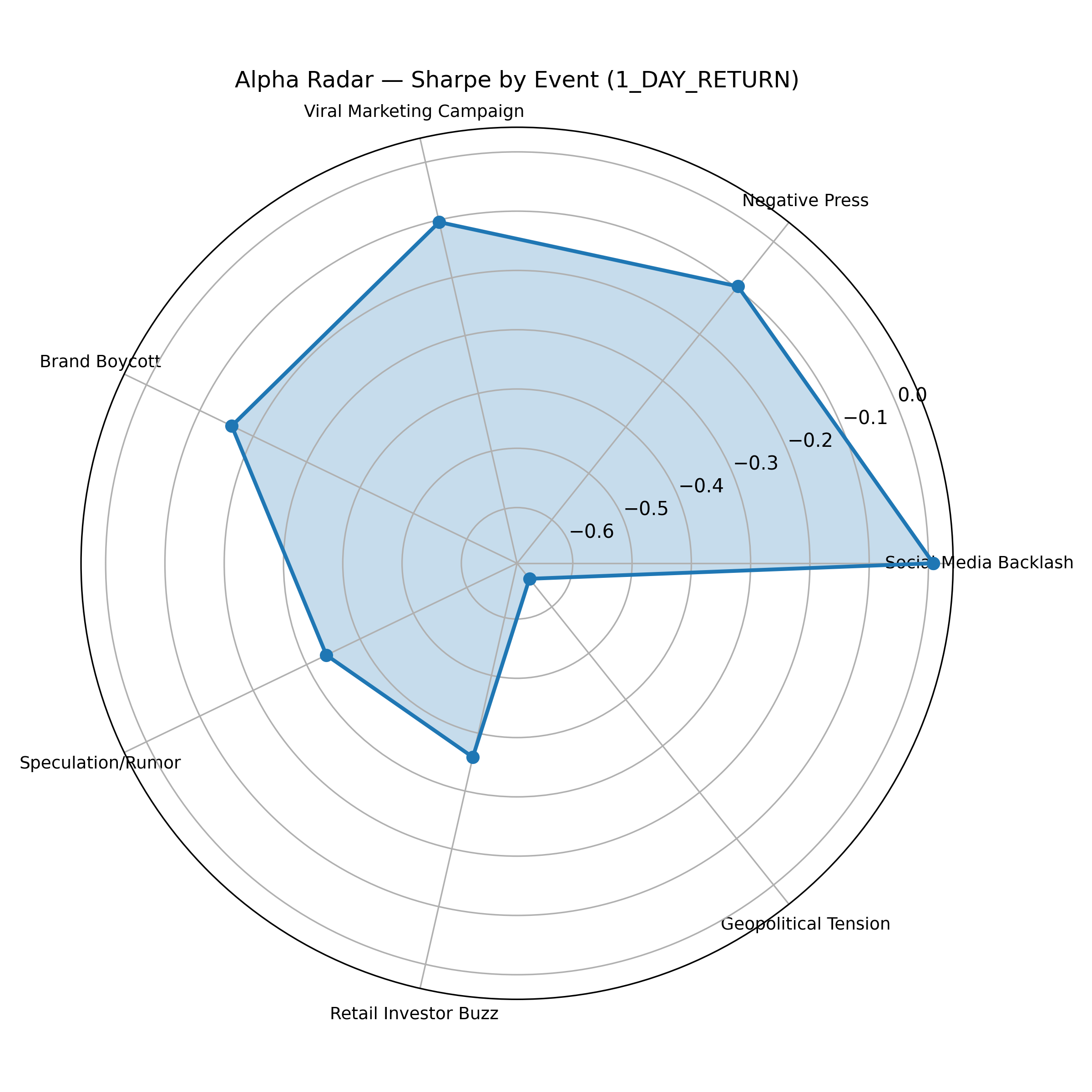} 
    \caption{Alpha radar for sharpe ratio of horizon one day.}
    \label{fig:single_panel_placeholder}
\end{figure}
\label{subsec:fig_placeholder_single}

\begin{figure}[h!]
    \centering
    \includegraphics[width=\linewidth]{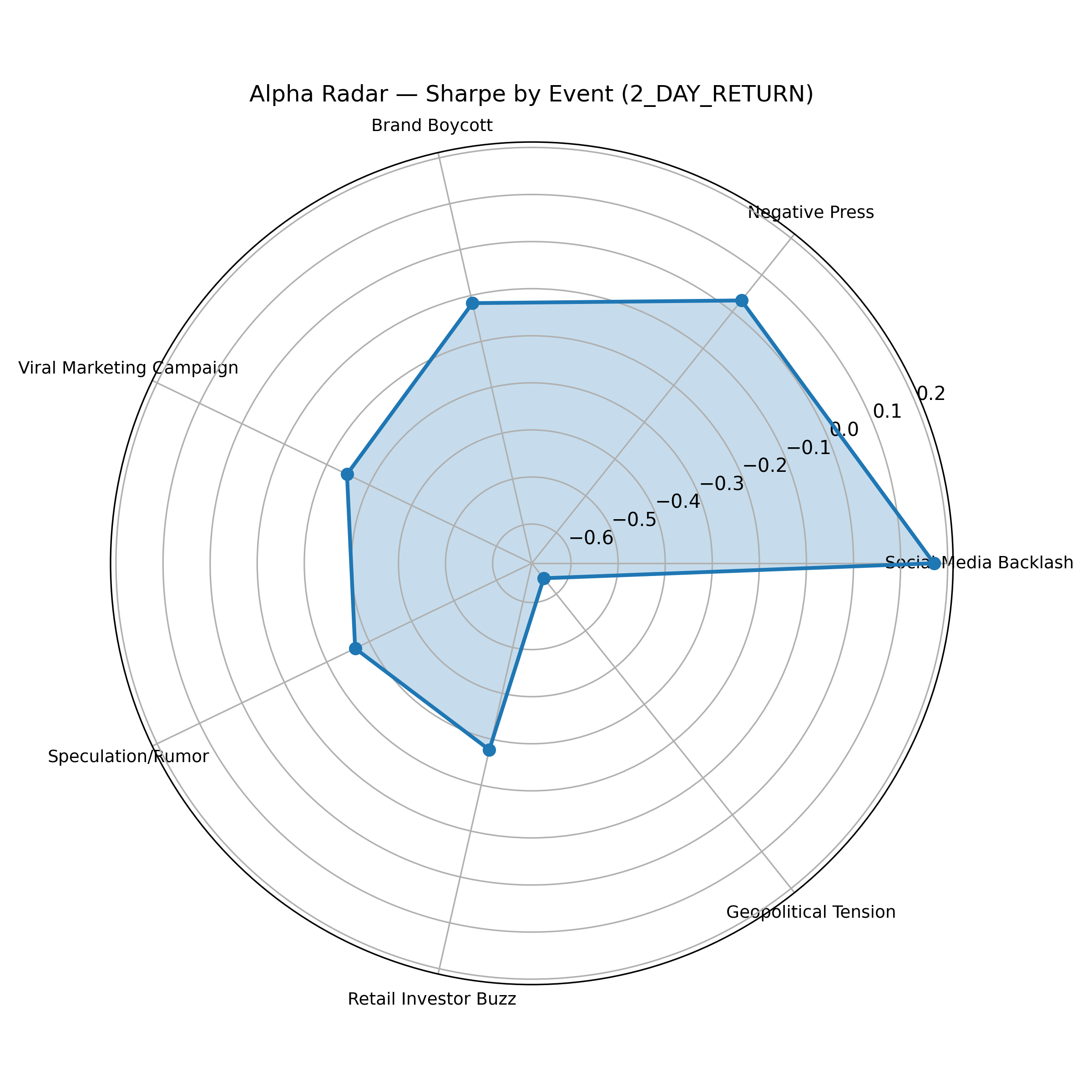} 
    \caption{Alpha radar for sharpe ratio of horizon two days.}
    \label{fig:single_panel_placeholder}
\end{figure}
\label{subsec:fig_placeholder_single}

\begin{figure}[h!]
    \centering
    \includegraphics[width=\linewidth]{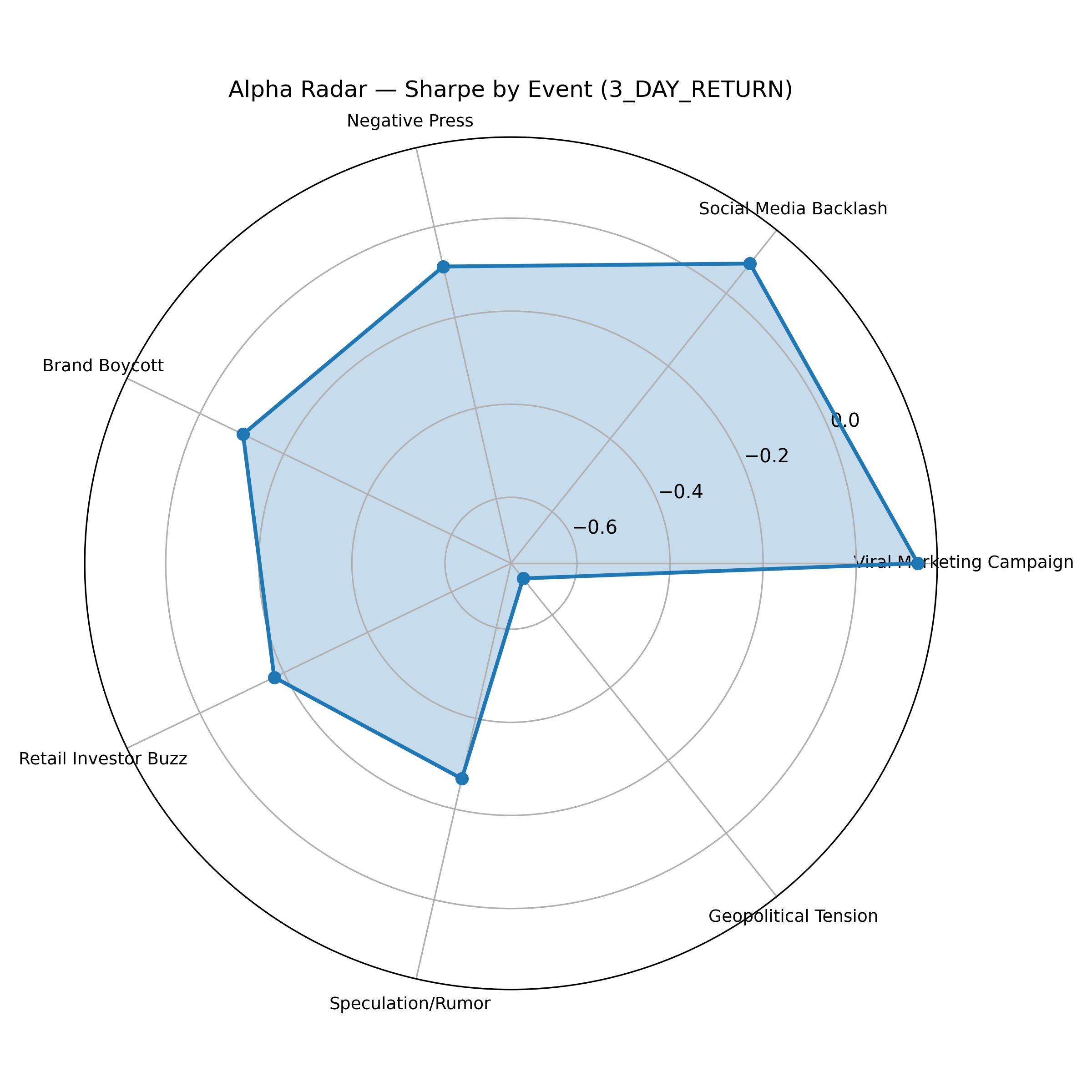} 
    \caption{Alpha radar for sharpe ratio of horizon three days.}
    \label{fig:single_panel_placeholder}
\end{figure}
\label{subsec:fig_placeholder_single}

\begin{figure}[h!]
    \centering
    \includegraphics[width=\linewidth]{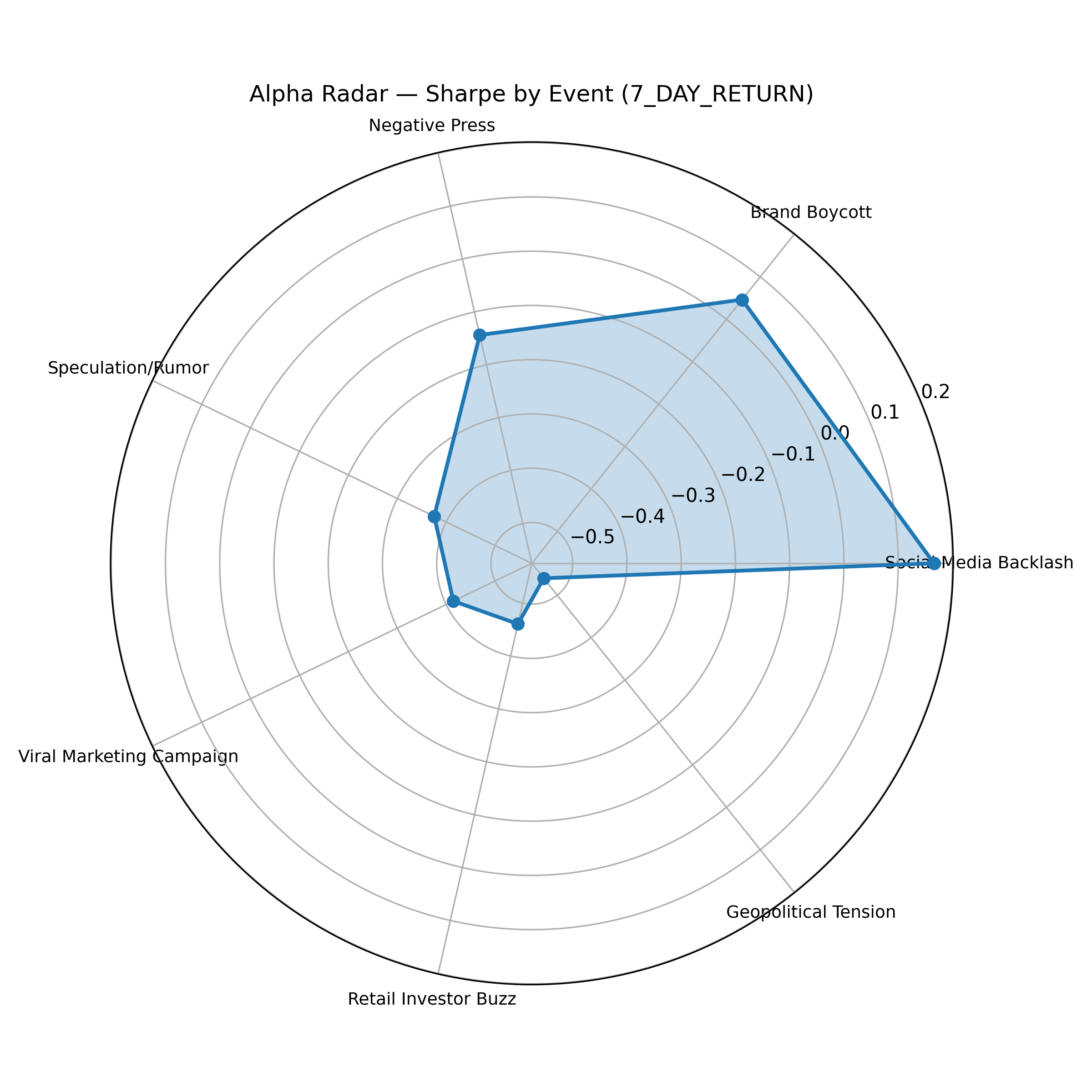} 
    \caption{Alpha radar for sharpe ratio of horizon seven days.}
    \label{fig:single_panel_placeholder}
\end{figure}
\label{subsec:fig_placeholder_single}

\subsection{Word Clouds}
\label{subsec:word_clouds}

\begin{figure}[h!]
    \centering
    \includegraphics[width=\linewidth]{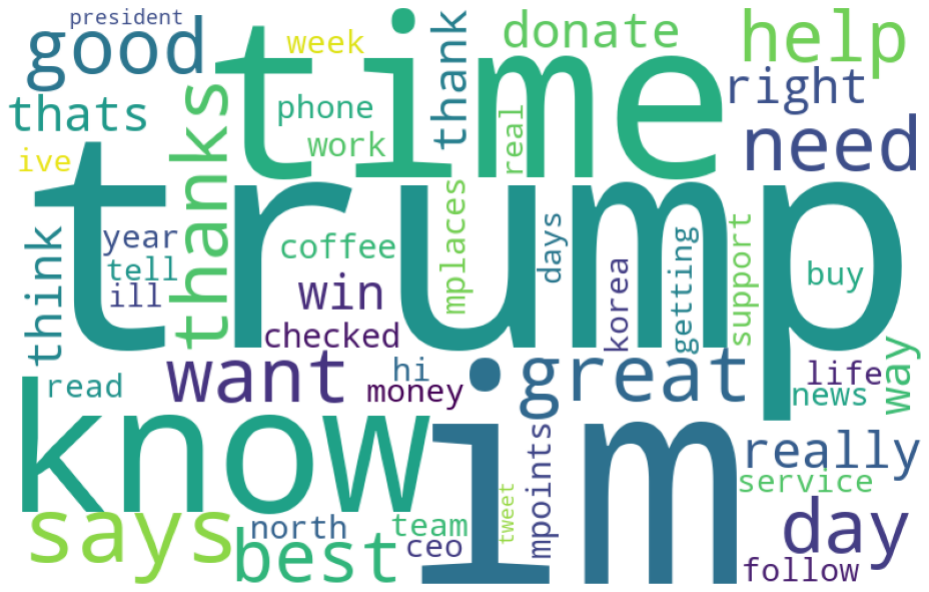} 
    \caption{Word cloud for LDA category: Political Discourse and Social Engagement.}
    \label{fig:single_panel_placeholder}
\end{figure}
\label{subsec:fig_placeholder_single}

\begin{figure}[h!]
    \centering
    \includegraphics[width=\linewidth]{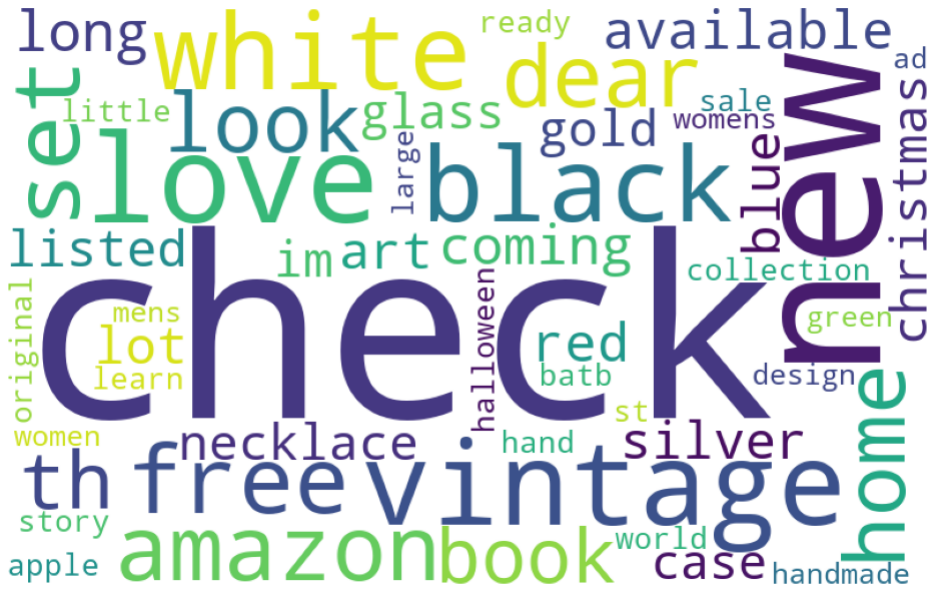} 
    \caption{Word cloud for LDA category: E-commerce Promotion and Lifestyle Products.}
    \label{fig:single_panel_placeholder}
\end{figure}
\label{subsec:fig_placeholder_single}

\end{document}